\documentclass[a4paper,11pt]{article}

\usepackage{jcappub} 
\usepackage{lineno}
\usepackage{longtable}
\usepackage{supertabular}
\usepackage{booktabs}
\usepackage{multirow}
\usepackage[utf8]{inputenc}   
\DeclareUnicodeCharacter{2212}{-}
\usepackage{hyperref}
\usepackage[T1]{fontenc}
\usepackage{appendix}
\usepackage{amsmath,amssymb}
\usepackage[makeroom]{cancel}
\usepackage{graphicx}
\usepackage{dcolumn}
\usepackage{booktabs}
\usepackage{multirow}
\usepackage{bm}
\usepackage{hyperref}
\usepackage{braket}
\usepackage{soul}
\usepackage{pgfplotstable}
\usepackage{booktabs} 

\usepackage{xcolor}     

\title{\boldmath Teukolsky by Design: A Hybrid Spectral-PINN solver for Kerr Quasinormal Modes}

\author[a,1]{Alexandre M. Pombo\note{Corresponding author.},}
\author[b]{Lorenzo Pizzuti}

\affiliation[a]{CEICO - Institute of Physics of the Czech Academy of Sciences}
\affiliation[b]{Dipartimento di Fisica “G. Occhialini”, Universit\`a degli Studi di Milano Bicocca, Piazza della Scienza 3, 20126 Milano, Italy}

\emailAdd{pombo@fzu.cz}
\emailAdd{lorenzo.pizzuti@unimib.it}

\abstract{We introduce \texttt{SpectralPINN}, a hybrid pseudo-spectral/physics-informed neural network (PINN) solver for Kerr quasinormal modes that targets the Teukolsky equation in both the separated (radial/angular) and joint two-dimensional formulations. The solver replaces standard neural activation functions with Chebyshev polynomials of the first kind and supports both soft -- via loss penalties -- and hard -- enforced by analytic masks -- implementations of Leaver's normalization. Benchmarking against Leaver’s continued-fraction method shows cumulative (real+imaginary part) relative frequency errors of $\sim 0.001\%$ for the separated formulation with hard normalization, $\sim 0.1\%$ for both the soft separated and soft joint formulations, and $\sim 0.01\%$ for the hard joint case. Exploiting our ability to solve the joint equation, we add a small quadrupolar perturbation to the Teukolsky operator, effectively rendering the problem non-separable. The resulting perturbed quasinormal modes are compared against the expected precision of the Einstein Telescope, allowing us to constrain the magnitude of the perturbation. These proof-of-concept results demonstrate that hybrid spectral-PINN solvers can provide a flexible pathway to quasinormal spectra in settings where separability, asymptotics, or field content become more intricate and high accuracy is required.}

\begin{document}

\maketitle

\flushbottom
%
\section{Introduction}\label{S1}
%
    The advent of gravitational-wave (GW) detections by the LIGO-Virgo-KAGRA (LVK) detector network~\cite{LVK_open_data_O4a} has opened a new window on the strong-field regime of gravity through spacetime dynamics. For the first time, we are able to probe the nature of compact objects (COs), that would otherwise be invisible, via their fingerprints in the GW emission from coalescing binaries. Since the first detection~\cite{abbott2016observation}, the LVK network has reported an ever-growing number of binary black hole (BH) and neutron star mergers, documented in the GWTC catalogs~\cite{GWTC1,GWTC2,GWTC3}. 

    Although general relativity (GR) is our most successful theory of gravity, it requires a dark sector (dark matter and dark energy) to match cosmological observations. In the strong-field regime, where direct tests remain limited, Kerr BHs are the canonical COs. At the same time, recent GW observations of BHs with component masses in or beyond the expected pair-instability gap (\textit{e.g.,}~\cite{AbbottGW190521,OBrianPISNGap,WoosleyPISNGap}) challenge standard stellar-evolution scenarios and motivate more complex formation channels, such as hierarchical mergers, as well as alternative CO models. It therefore remains an open question how robustly current and future GW observations can distinguish GR Kerr BHs from other possible COs.

    Assessing the viability of alternative COs requires analyzing their oscillations through quasi-normal modes (QNMs). When perturbed, a CO oscillates in a discrete set of characteristic QNMs, whose complex frequencies and mode amplitudes act as fingerprints and stability probes. As gravitational systems, COs radiate part of their oscillation energy as GWs, both after external perturbations and in the post-merger ringdown phase. 

    So far, the dominant $(\ell,m,n) = (2,2,0)$ mode has been directly measured in LVK data~\cite{abbott2016observation}, but the detection of sub-dominant modes or overtones remains challenging because of the limited signal-to-noise ratio (SNR) of current detectors. Several studies indicate that full, mode-resolved spectroscopic tests of GR require ringdown SNRs of $\mathcal{O}(100)$ or higher~\cite{Bhagwat2020}. Such SNRs are expected, for favorable sources, with next-generation observatories -- the Einstein Telescope (ET)~\cite{Punturo2010ET} and Cosmic Explorer (CE)~\cite{Evans2021CE,Hall2022CE} -- and with space-based detectors such as LISA~\cite{LISA2017}, which should be able to resolve multiple $\big($typically $\sim 3$: (2,2,0), (3,3,0), (2,0,0)$\big)$ individual modes with high precision. The next generation of detectors will thus provide ringdown data of unmatched quality, creating a unique opportunity to test whether observed remnants are consistent with Kerr BHs or with alternative CO scenarios. Delivering accurate QNM predictions across these possibilities is therefore essential, yet remains a major theoretical and computational challenge.
    
    From a theoretical standpoint, the ringdown of a rotating BH in GR is described by linear perturbations of the Kerr spacetime. In Boyer-Lindquist coordinates, these perturbations are encoded in a single complex master field $\psi$ of spin weight $s$ that obeys the Teukolsky master equation~\cite{teukolsky1973perturbations}. The standard treatment exploits separability into radial and angular components, leading to two coupled ordinary differential equations. Solving these equations with appropriate boundary conditions selects a discrete set of complex frequencies $\omega_{s\ell m n}$: the Kerr QNMs (\textit{e.g.,}~\cite{Berti2006,Abbott21}). These Kerr QNMs serve as the baseline for ringdown spectroscopy and as a reference against which modifications induced by alternative COs, extra fields (\textit{e.g.,}~\cite{Isi2021}), or deviations from GR can be quantified. Hence, realizing this spectroscopic potential requires accurate and flexible theoretical predictions for QNMs, both for Kerr BHs and for possible deviations associated with alternative COs or modified gravity.

    A variety of numerical strategies have been developed to extract these frequencies. One of the most widely used is Leaver's continued-fraction method~\cite{leaver1985analytic}, which is based on Frobenius series expansions and three-term recurrence/quotient relations. It yields extremely accurate spectra for separable problems with known asymptotics and has become a benchmark for Kerr QNMs. However, methods that rely so heavily on separability and continued fractions become increasingly cumbersome, or even inapplicable, when the QNM equation is no longer exactly separable, when the asymptotic structure is modified, or when extra degrees of freedom alter the boundary conditions. This motivates the development of solvers that treat the full partial differential equation (PDE) in the radial and polar coordinates $(r,\theta)$ and can be adapted to non-Kerr or non-separable settings with minimal redesign.

    Recent progress in this direction includes spectral-decomposition schemes, in particular the pseudo-spectral method (PSM) developed by \cite{Blazquez2024}, which expands the perturbation in Chebyshev polynomials and associated-Legendre functions over a compactified domain. Such methods can resolve large portions of the Kerr QNM spectrum -- including fundamental modes and overtones, and rapidly rotating BHs -- with fractional errors between $10^{-6}$ and $10^{-3}$ depending on spin and excitation order~\cite{Blazquez2024}. In parallel, physics-informed neural networks (PINNs) have emerged as an alternative strategy in which the differential equation is built directly into the loss function and the QNM frequencies are treated as trainable parameters. A first implementation for the Kerr problem achieved sub-percent accuracy with respect to the continued-fraction method across a broad range of spin parameters and mode families~\cite{luna2023solving}. Beyond their flexibility, PINNs do not require explicit numerical differentiation schemes or fixed meshes, and they naturally accommodate systems with multiple coupled PDEs and several eigenvalue parameters.
    
    Taken together, these developments suggest that hybrid approaches combining the structured accuracy of spectral solvers with the flexibility of neural operators are particularly promising for QNM spectroscopy and CO discrimination. In this work, we build on this idea and introduce \texttt{SpectralPINN}, a hybrid PSM/PINN solver tailored to the Teukolsky equation in both its separated (radial/angular) and its full two-dimensional (2D) $(r,\theta)$ formulation.
    
    The central feature of our solver is the substitution of standard neural activation functions with Chebyshev polynomials of the first kind, enabling the network to act simultaneously as a spectral interpolator and as a PINN enforcing the PDE residual and eigenvalue conditions. Boundary/normalization conditions can be imposed through hard analytic masks encoding the horizon and infinity asymptotic behaviors (\textit{aka} hard enforcement), or via additional loss penalties (\textit{aka} soft enforcement). We compare the resulting $(\mathbb{R}+\mathbb{I})$ QNM frequencies with the results of the Leaver's continued-fraction method for the separated problem with hard enforcement, for separated soft enforcement runs, and for the full 2D Teukolsky PDE.
    
    Exploiting the joint $(r,\theta)$ formulation, we then consider a controlled quadrupolar perturbation of the Teukolsky operator that breaks exact separability while preserving the asymptotic structure. Starting from the Kerr solution, we track the fundamental gravitational $s=-2,\ell=2,m=2$ and $m=0$ QNM as a function of the perturbation amplitude, compute the induced shifts in frequency and damping time, and map them into time-domain ringdown signals. These perturbed ringdowns are compared with their Kerr counterparts using the ET sensitivity curve, allowing us to estimate the minimum ringdown SNR required to resolve a given perturbation and to forecast the constraints that next-generation detectors can place on such deviations.
    
    With simultaneously high precision and robust accuracy in both separable and mildly non-separable full-PDE regimes, \texttt{SpectralPINN} provides a unified computational pathway to QNM spectra in contexts where separability, asymptotics, or field content become more intricate -- for example in rapidly rotating near-extremal BHs, matter-coupled systems, and perturbations in modified-gravity models. This makes \texttt{SpectralPINN} a natural building block for the flexible, high-accuracy QNM modeling that will be required to fully exploit the discriminating power of next-generation ringdown observations for distinguishing between different classes of COs.

    The paper is organized as follows. Sec.~\ref{S2} reviews the Teukolsky equation, Secs.~\ref{S3}-\ref{S4} introduce the \texttt{SpectralPINN} solver and describe, Sec.~\ref{S5} applies it to a deformed Kerr ringdown and ET forecasts.

    Unless otherwise stated, throughout the text we consider geometrid units where $G=c= 1$, with $G$ the Newton's constant and $c$ denotes the speed of light. The signature of the spacetime is always $(-,+,+,+)$. In the case of a single dependence function, we always denote the the derivative in order to the dependence coordinate as a prime, $dZ(z)/dz=Z'(z)$.
%
\section{Teukolsky equation}\label{S2}
%
    The rotating vacuum BH solution of the Einstein equations is described by the Kerr metric. In Boyer-Lindquist ($t,r, \theta, \varphi$) coordinates, this is given by
    \begin{equation}
        ds^2 = -\bigg(1-\frac{2M r}{\Sigma}\bigg)dt^2-\frac{4M a r \sin ^2 \theta}{\Sigma}dt d\varphi+\Sigma\Big(\frac{dr^2}{\Delta}+ d\theta ^2\Big)+\bigg( r^2 + a^2+\frac{2M a^2 r \sin ^2 \theta}{\Sigma}\bigg)\sin ^2\theta d\varphi ^2\ ,
    \end{equation}
    with
    \begin{equation}
        \Delta = r^2-2Mr+a^2\ , \qquad \qquad \Sigma = r^2+a^2\cos ^2 \theta \ ,
    \end{equation}
    and $M$ the mass of the Kerr BH and $0\leqslant a \leqslant M$ the rotation parameter of the Kerr metric with intrinsic angular momentum $J = M a$. The zeros of the $\Delta$ function determine the inner and outer horizon positions
    \begin{equation}
        r_\pm = M\pm \sqrt{M^2-a^2}\ .
    \end{equation}
    The linear perturbations of a Kerr BH metric are encoded in the Teukolsky~\cite{teukolsky1973perturbations} master equation as a function of the master variable $\psi (t,r,\theta,\varphi)$, with $s=0$ denoting a scalar, $s=-1$ a vector (electromagnetic), and $s=-2$ a tensor (gravitational) perturbation. Following \cite{teukolsky1973perturbations,kokkotas1999quasi,ishak2018gravitational}, the standard Teukolsky master equation can be written as,
    \begin{align}
     &\Bigg[\frac{\big(r^2+a^2\big)^2}{\Delta}-a^2 \sin ^2 \theta\Bigg]\frac{\partial ^2 \psi}{\partial t^2}+\frac{4 M a r}{\Delta}\frac{\partial ^2 \psi}{\partial t \partial \varphi}+\Bigg[\frac{a^2}{\Delta}-\frac{1}{\sin ^2\theta}\Bigg]\frac{\partial ^2 \psi}{\partial \varphi ^2}\nonumber\\
     &-\Delta ^{-s}\frac{\partial }{\partial r}\bigg( \Delta^{s+1}\frac{\partial \psi}{\partial r}\bigg)-\frac{1}{\sin \theta} \frac{\partial }{\partial \theta}\bigg(\sin \theta \frac{\partial \psi}{\partial \theta}\bigg)-2s \bigg[\frac{a (r-M)}{\Delta}+\frac{i \cos \theta}{\sin ^2\theta}\bigg]\frac{\partial \psi}{\partial \varphi}\\
     &-2s \Bigg[\frac{M(r^2-a^2)}{\Delta}-r-i a \cos \theta \Bigg]\frac{\partial \psi}{\partial t}+\big(s^2\cot ^2 \theta -s\big) \psi = 0\ .\nonumber
    \end{align}
    In this work we will solely consider the gravitational perturbations associated with the tensor $s=-2$ mode. Following the standard procedure (\textit{e.g.,} \cite{leaver1985analytic,luna2023solving,Blazquez2024}), the perturbation function, $\psi$, is assumed to possess an harmonic time behavior $\psi (t,r,\theta,\varphi)  \equiv e^{-i \omega t} \psi (r,\theta, \varphi)$, with $\omega$ the (complex) frequency of the mode. Teukolsky found that the function admits a further separation into a product of a purely radial, $r$, and zenithal angular, $\theta$, coordinates
    \begin{equation}\label{E2.5}
     \psi = \frac{1}{2\pi} \sum_{\ell =|s|} ^{+\infty}\, \sum_{m=-\ell} ^{\ell} e^{i m \varphi} R_\ell ^m (r) S_{\ell}^m(\theta) \ ,
    \end{equation}
    where $m$ is the azimuthal quantum number. Note that, in contrast to the case of standard spherical harmonics, the separation is $\omega-$dependent: $R_\ell ^m(r)\equiv R_\ell ^m(r,\omega)$. Hereafter, the $\ell,\ m$ indices will be omitted when not needed, however it is important to point out that the functions and characteristic parameters are all $s,\, \ell$ and $m$ dependent. The result are two ordinary differential equations (ODEs), one for the radial and one for the angular parts of the total solution. When solving for the modes, we take $2M=1$. The radial part of the solution has the form,
    \begin{equation}\label{E2.6}
        \Delta R'' (r)+ (s+1)(2\,r-1) R'(r) + V(r) R(r) = \Lambda\, R(r)\ ,
    \end{equation}
    where $\Lambda \equiv \Lambda _\ell ^m$ is a separation constant that depends on the angular momentum numbers $(\ell,\,m)$ and the frequency $\omega$. In the Schwarzschild limit ($a\to0$), it reduces to~\cite{leaver1985analytic}
    \begin{equation}\label{E2.7}
        \lim _{a\to 0} \Lambda = \ell(\ell+1)-s(s+1)\ .
    \end{equation}
    The potential term function, $V(r)$, is given as,
    \begin{equation}
        V(r) = \bigg[\frac{\big(r^2+a^2\big) ^2 }{\Delta }-a^2\bigg]\omega ^2+\bigg[\frac{i s \big(a^2-r^2 \big)-2amr }{\Delta }+2isr\bigg]\omega+\frac{a^2 m^2+isam(2r-1)}{\Delta}\ .
    \end{equation}
    The angular equation reads,
    \begin{equation}\label{E2.9}
        (1-y^2)S''(y)-2y S'(y)+\bigg[a^2 \omega ^2 y^2-2 a \omega s y+s -\frac{(m+ys)^2}{1-y^2}\bigg]S(y)=-\Lambda S(y)\ ,
    \end{equation}
    where we already performed the coordinate compactification $y= \cos \theta$. Physical behavior is imposed through ingoing boundary conditions at the horizon, $r_+$, and outgoing at spatial infinity, while requiring regularity at the poles, $\theta = 0,\, \pi$. Following Leaver~\cite{leaver1985analytic}, we consider the following ans\"atze,
    \begin{align}\label{E2.10}
        &R(r) = e^{i\omega r}(r-r_-)^{\sigma_-}(r-r_+)^{\sigma_+}f(x)\ ,\nonumber\\
        \sigma_- = i \omega -1&-s+i\frac{\omega r_+-am}{\sqrt{1-4a^2}}\ ,\qquad\qquad \sigma_+= -s-i\frac{\omega r_+-am}{\sqrt{1-4a^2}}\ .
    \end{align}
    where the radial coordinate has been compactified as $x=1/r$, such that the resulting radial domain $r\in [r_+, +\infty[$, becomes $x\in [0,1]$. The angular function is transformed as,
    \begin{equation}\label{E2.11}
        S(y) = e^{a\omega y}(1+y)^{\frac{|m-s|}{2}}(1-y)^{\frac{|m+s|}{2}}g(y)\ .
    \end{equation}
    The functions $f(x)$ and $g(y)$ have to be regular in the compactified domains $[0,1]$ and $[-1,1]$, respectively. By replacing \eqref{E2.10} into \eqref{E2.6} and \eqref{E2.11} into \eqref{E2.9}, we obtain the set of two coupled (through the separation constant, $\Lambda$) ODEs for $f(x)$ and $g(y)$ of the form
    \begin{align}\label{E2.12}
        \mathcal{L}_F&=F_{xx}\, f''(x)+F_x\, f'(x)+F_0\, f(x)+F_\Lambda\, f(x) = 0\ ,\nonumber\\
        \mathcal{L}_G&=G_{yy}\, g''(y)+G_y\, g'(y)+G_0\, g(y)+G_\Lambda\, g(y) = 0\ .
    \end{align}
    Following the same convention as \cite{luna2023solving}, the explicit forms of the $F_i$ and $G_i$ functions are given in Appendix~\ref{A2}. With this, the problem reduces to finding the set of $\{\omega,\, \Lambda\}$ that satisfy \eqref{E2.12} for a given set of $a/M,\, \ell$ and $m$. 
    
    While the standard procedure relies on the computation of the two sets of coupled ODEs, it is also possible to solve the Teukolsky equation through a single 2D function $\phi(x,y)=f(x)\cdot g(y)$. The single 2D PDE is obtained either by replacing $\Lambda$ in one of the equations by the one of the other -- after solving the second in order to $\Lambda$ --, or by considering \eqref{E2.5} without the separation of coordinates, such that
    \begin{equation}
        \psi = \frac{1}{2\pi} \sum_{\ell=|s|} ^{+\infty}\, \sum_{m=-\ell} ^{\ell} e^{i m \varphi} \phi _\ell ^m (r,\,\theta) \ .
    \end{equation}
    Performing the same coordinate compactification and imposing the correct physical behavior through the proper boundary conditions,
    \begin{equation}
        \phi(r,y) = e^{\omega (a y+i r)}(1+y)^{\frac{|m-s|}{2}}(1-y)^{\frac{|m+s|}{2}}(r-r_-)^{\sigma_-}(r-r_+)^{\sigma_+}p(x,y)\ .
    \end{equation}   
    The resulting single 2D PDE can then be cast as
    \begin{equation}\label{E2.16}
        \mathcal{L}_P = P_0\, p+ P_x\,\partial_x p+P_{xx} \,\partial_{xx} p+P_y\,\partial_y p+ P_{yy}\,\partial_{yy} p = 0\ ,
    \end{equation}
    with the $P_{ij}$ functions given explicitly in Appendix~\ref{A2}. Observe that by solving the full PDE \eqref{E2.16} one loses the need to fit the separation constant, $\Lambda$, and the problem reduces to finding the frequencies $\omega$ that satisfy \eqref{E2.16}. As a recall, the functions $F_i$ are explicitly dependent on $x$, the $G_i$ functions on $y$ and the $P_{ij}$ functions depend on both $x$ and $y$. The dependence was here omitted for notation simplicity\footnote{We stress that our numerical methods, in particular \texttt{SpectralPINN}, are sensitive to sharp features in the integrands. For this reason the $F_i$, $G_i$ and $P_{ij}$ have been algebraically simplified to remove spurious divergent terms before being passed to the solver.}.
%
\section{Neural Networks}\label{S3}
%
    In this section we introduce the neural network solvers employed in the solution of the Teukolsky equation for the Kerr QNMs.
%
    \subsection{PINNs}\label{S3.1}
%
    As a baseline for our hybrid solver, we first construct a standard PINN following \cite{cornell2024solving,luna2023solving} that solves the separated system~\eqref{E2.12}. To enforce the radial and angular equations, we introduce two one-dimensional grids in the compactified coordinates $x$ and $y$. For simplicity, and because the PINN is not particularly sensitive to the exact collocation pattern, we use uniformly spaced grids
    \begin{align}
        x_i &= \frac{i-1}{N_x-1}\ , \qquad\qquad\, i = 1, ..., N_x\ ,\\
        y_j &= 2\frac{j-1}{N_x-1}-1\ , \qquad j=1,...,N_y\ ,
    \end{align}
    with $N_x$ and $N_y$ the number of grid points. In all the evaluations (including the hybrid method, see Sec.~\ref{S3.2}), we consider $N_x=N_y = 100$. For the loss function, we consider the weighted sum of the average moduli, $L_1$, of the right-hand side of \eqref{E2.12},
    \begin{equation}
        \mathcal{L} = \frac{W}{N_x} \sum_{i=0} ^{N_x}\big|\mathcal{L}_F\big|+\frac{1}{N_x} \sum_{i=0} ^{N_x}\big|\mathcal{L}_G\big|+\beta \mathcal{L}_{\rm Soft}\ ,
    \end{equation}
    where the factor $W>1$ enhances the contribution of the radial residual (more intricate) relative to the angular one, and $\beta$ enhances the soft normalization residual in relation to the bulk (radial+angular). We verified that choosing $W$ in the range $\mathcal{O}(1\text{--}20)$ produces very similar eigenvalues; in all results shown we simply fix $W=10$ following \cite{luna2023solving}. By contrast, the value of $\beta$ has a strong impact on the converged eigenvalues $\{\omega,\Lambda\}$. $\beta$ is varied depending on the specific soft normalization adopted.
    
    To solve the two ODEs~\eqref{E2.12}, we approximate the unknown functions $f(x)$ and $g(y)$ with two independent fully connected networks, $\mathcal{F}$ and $\mathcal{G}$. Each network takes a single scalar input ($x$ for $\mathcal{F}$ and $y$ for $\mathcal{G}$) and consists of four hidden layers with $40$–$60$–$60$–$40$ neurons. This architecture provided the best accuracy among the configurations we tested. The output layer of each network has two neurons, representing the real and imaginary parts of $f$ or $g$, so that the networks remain purely real-valued while approximating complex functions. The trainable parameters in the standard PINN configuration are therefore the weights and biases of both networks together with the complex eigen-parameters $\{\omega,\Lambda\}$ (see Fig. 1 of ~\cite{luna2023solving} for a scheme of the architecture).

    Derivatives of $f$ and $g$ with respect to their inputs are computed via automatic differentiation using \texttt{PyTorch}, which makes the implementation straightforward but also introduces a small amount of numerical noise in the residuals. The gradients of the loss with respect to all network parameters, as well as to $\omega$ and $\Lambda$, are obtained through the same automatic differentiation mechanism, so that both networks and the eigenvalues are trained jointly as a single model. The optimization algorithm and learning-rate schedule will be described in Sec.~\ref{S4}.

    To obtain QNMs over a range of spins $a/M \in [0,1[$ we adopt a sequential (continuation) strategy in $a$, which will later also be used for \texttt{SpectralPINN}. We start with the Schwarzschild case ($a=0$), with the weights of both networks initialized from Gaussian distributions with zero mean and standard deviations $\sigma_f=0.05$ for $\mathcal{F}$ and $\sigma_g=0.01$ for $\mathcal{G}$, and all biases set to zero. The initial value of $\omega$ at $a=0$ is taken from the known Schwarzschild QNM for the chosen $(\ell,m)$, while the separation constant is initialized to its analytic Schwarzschild value \eqref{E2.7}.

    Once a converged solution is obtained for $a=0$, we increase the spin parameter in small steps and retrain the networks at each new value of $a$, using as initial conditions the weights, biases and eigenvalues from the previous step. We repeat this procedure until we reach the maximum spin $a=0.4999$. In this way, convergence at each new spin starts from a configuration that is already close to the target mode, rather than from a random initialization. For the standard PINN this mainly improves convergence speed and stability (see \cite{luna2023solving}); for \texttt{SpectralPINN}, which is more sensitive to the initial amplitudes, it is crucial to avoid convergence to spurious local minima or to modes with different $(\ell,m)$. The sequence of spin values is chosen with a denser sampling at higher spins, where the QNMs vary more rapidly and the equations become more challenging to solve.
%
    \subsection{SpectralPINN}\label{S3.2}
%
    To numerically solve the coupled ODEs~\eqref{E2.12} and the single 2D PDE~\eqref{E2.16} describing the perturbation functions, we use an in-house solver, \texttt{SpectralPINN}, which combines a PSM, inspired by \cite{Blazquez2024}, with a PINN-like training strategy, inspired by \cite{luna2023solving}. The key idea is to represent the unknown functions in a Chebyshev basis and to treat the corresponding spectral amplitudes as trainable parameters in a neural network optimization loop.

    For the separated problem, we expand the radial and angular functions as
    \begin{equation}\label{E3.1}
        f(x) = \sum_{i=0}^{N} \mathcal{A}^f_{i}\,T_i(x)\ , \qquad g(y) = \sum_{j=0}^{L} \mathcal{A}^g_{j}\,T_j(y)\ ,
    \end{equation}
    and for the joint formulation we expand the 2D function $p(x,y)$ as
    \begin{equation}\label{E3.2}
        p(x,y) = \sum_{i=0}^{N}\sum_{j=0}^{L} \mathcal{A}^p_{ij}\,T_i(x)\,T_j(y)\ ,
    \end{equation}
    where $T_n$ denotes the Chebyshev polynomials of the first kind, and $N,\, L$ are the maximum orders retained in the radial and angular directions, respectively. Because of the compactification used in Sec.~\ref{S2}, the angular dependence has already been mapped to $y\in[-1,1]$ and the explicit spin-weighted spherical-harmonic structure is encoded in the pre-factors. In this setting we found that expanding also the angular dependence in Chebyshev polynomials leads to a more stable and accurate solver than using associated-Legendre polynomials directly like the strategy used in \cite{pombo2025dark}.
    
    In a standard PSM, inserting the expansions~\eqref{E3.1} and~\eqref{E3.2} into the field equations and evaluating them on a collocation grid yields a (non-)linear algebraic system for the coefficients $\mathcal{A}$. Solving this system gives the spectral amplitudes, with exponential convergence in $N,\, L$ for smooth solutions. Alternatively, one can view \eqref{E3.1} and~\eqref{E3.2} as parameterizations of the solution and determine the coefficients $\mathcal{A}$ by minimizing the residuals of \eqref{E2.12} and~\eqref{E2.16}, together with the imposed boundary/normalization conditions and any additional physical constraints (see Sec.~\ref{S4.3}). This residual-minimization viewpoint is naturally aligned with the PINN framework.
  
\begin{figure}[h!]
	   \centering
	   \includegraphics[width=0.95\textwidth]{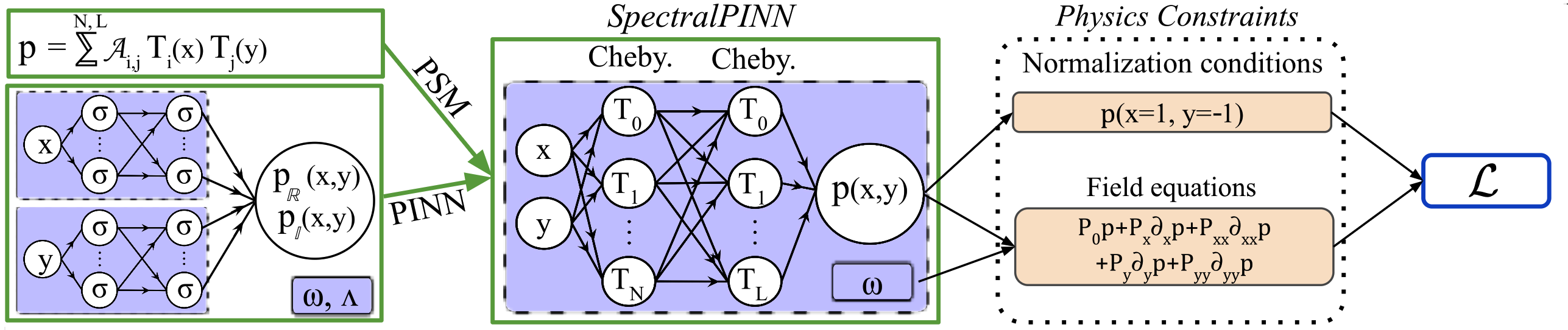}\\
	   \caption{Schematic representation of the \texttt{SpectralPINN} architecture. Each neuron corresponds to a Chebyshev basis function, and each ``layer'' is associated with one of the coordinate dependence. The scheme is loosely inspired by the PINN designs in \cite{luna2023solving}.}
	   \label{F1}
    \end{figure}

    Our \texttt{SpectralPINN} solver is precisely such a hybrid: it keeps the PINN-style training loop (gradient-based minimization of a residual-based loss) but replaces generic non-linear activations with \emph{spectral activations} that enforce the problem’s symmetry. In practice, the ``layers'' of the network correspond to the coordinate directions, and the ``neurons'' are the basis functions $T_i(x)$ and $T_j(y)$. The only learnable parameters are the spectral amplitudes $\mathcal{A}^f_i$ and $\mathcal{A}^g_j$ or $\mathcal{A}^p_{i,j}$, together with the eigenvalues $(\omega,\Lambda)$. A schematic representation of this architecture is shown in Fig.~\ref{F1}. This design sacrifices some of the flexibility of generic PINNs, but gains efficiency and a strong inductive bias towards solutions compatible with the assumed symmetry and basis.

    Following standard spectral practice, we evaluate the basis functions on collocation grids matched to their orthogonality properties. For the compactified radial coordinate $x\in[0,1]$ we use a mapped Chebyshev-Gauss-Lobatto grid, and for the angular coordinate $y\in [-1,1]$ we use the usual Gauss-Lobatto nodes:
    \begin{align}\label{E3.3}
      x_i &=\frac{1}{2}\bigg[1-\cos\!\left(\frac{i\pi}{N_x}\right)\bigg]\ ,\qquad i=0,\, \dots\, ,N_x\ ,\nonumber\\
      y_j &=\cos\!\left(\frac{j\pi}{N_y}\right)\ ,\qquad \qquad \quad\ \,j=0,\, \dots\, ,N_y\ ,
    \end{align}
    with $N_x\geqslant N$ and $N_y\geqslant L$ the numbers of collocation points in the radial and angular directions, respectively. The grid points are fixed \textit{a priori} and depend only on $(N_x,\, N_y)$.

    Since the evaluation points are known in advance and the basis functions depend only on these grid points, we can pre-compute: \textit{i)} the Chebyshev values $T_n(x_k)$ and $T_n(y_q)$ and the corresponding differentiation matrices $D_x$, $D_{xx}$ (first and second derivatives in $x$) and $D_y$, $D_{yy}$ (first and second derivatives in $y$), obtained from the standard recurrence relations for Chebyshev polynomials~\cite{MasonHandscomb2003};  \textit{ii}) all the coordinate-dependent coefficient functions appearing in the compactified functions \eqref{E2.12} and \eqref{E2.16}, \textit{i.e.,}\ $\mathcal{F}_i(x_k)$, $\mathcal{G}_i(y_q)$ and $\mathcal{P}_i(x_k,y_q)$. Because the derivatives of the basis are known analytically, no additional numerical differentiation scheme is required, which simplifies the implementation and improves accuracy. These pre-computations also avoid redundant calculations at each training iteration and significantly reduce the computational cost.
    In each training cycle, the current set of amplitudes and eigenvalues is used to reconstruct the functions $f$, $g$ or $p$ on the collocation grid, together with their derivatives, via the pre-computed differentiation matrices. These reconstructions are inserted into the operators $\mathcal{L}_F$, $\mathcal{L}_G$ and $\mathcal{L}_P$ to obtain the residuals of the radial, angular and 2D equations. The residuals then enter the loss functional as the main contribution. Additional terms in the loss enforce soft normalization and (when required) other physical conditions; the detailed loss decomposition is discussed in Sec.~\ref{S4.3}.

    From a machine-learning perspective, \texttt{SpectralPINN} is just a neural network where the neurons are basis functions, the weights are the spectral amplitudes, the target output is a vanishing residual for the governing ODEs/PDE and the imposed boundary/normalization conditions. Unless otherwise stated, the solver, as well as the PINNs, uses a mean absolute error (MAE) loss on the residuals,
    \begin{equation}
        \mathrm{MAE} = \frac{1}{n}\sum_{i=1}^n \big| k_i \big|\ ,
    \end{equation}
    where $k_i$ denotes the residual evaluated at the $i^{th}$ collocation point. We have tested mean squared error (MSE, \textit{aka $L_2$}) losses as well, but given that our pre-training strategy already yields relatively small initial residuals, MAE tends to focus more effectively on reducing the remaining small deviations. The optimization and learning-rate schedule are described in the next section.
    \begin{table}
     \centering
     \caption{Percentual cumulative ($\mathbb{R}+\mathbb{I}$) relative error of the perturbation frequency $\omega$ computed with Leaver’s continued-fraction value for $(s,l,m)=(-2,2,0)$ and $a/M=0.90$ as a function of the number of basis terms $N\times L$ for $N_x\times N_y = 100\times 100$. The computation is performed with a hard normalization (see Sec.~\ref{S5}) \texttt{complex128} type, \texttt{HAdamD} optimizer and separated \texttt{ReducedLROnPlateau} scheduler (see Sec.~\ref{S4}).}
     \vspace{2mm}
     \label{T1}
        \begin{tabular}{c|ccccccc|}
         \cline{2-8}
         \multicolumn{1}{c}{} & \multicolumn{7}{|c|}{$N \times L$} \\
         \cline{2-8}
         Eq. type & $10\times 10$ & $15\times 10$ & $10\times 15$ & $15\times 15$ & $20 \times 20$ & $30 \times 30$ & $50\times 50$  \\
         \hline
         ODE & $0.714$ & $0.109$ & $0.156$ & $0.012$ & $0.006$ & $0.002$ & $0.429$\\
         PDE & $0.920$ & $0.234$ & $0.280$ & $0.071$ & $0.052$ & $0.011$ & $0.600$ \\
         \hline
        \end{tabular}
    \end{table}

    Table~\ref{T1} illustrates the convergence of \texttt{SpectralPINN} as the number of Chebyshev modes $N\times L$ is increased. For both the
    separated ODE system and the full 2D PDE, the cumulative $(\mathbb{R}+\mathbb{I})$ relative error with respect to Leaver’s frequency decreases rapidly when going from $10\times 10$ to $30\times 30$, reaching the $\sim 10^{-3}\%$ level for the ODE case and $\sim 10^{-2}\%$ for the PDE. Pushing the basis to $50\times 50$ slightly degrades the accuracy, likely reflecting a combination of ill-conditioning, numerical noise, and the increased difficulty of fully training the larger spectral network. In practice, moderate spectral resolutions provide an optimal balance between accuracy and trainability for the Kerr QNMs considered here; in all following computations, we take $N\times L = 30\times 30$.

    For illustration, Table~\ref{tab:weights_a045_6x6} in Appendix~\ref{A1} lists the first $6\times 6$ Chebyshev coefficients $\mathcal{A}_{ij}$ for the $(s,\ell,m)=(-2,2,0)$ mode at $a/M=0.90$. The pattern shows sizable low-order coefficients and rapidly decaying higher-order entries, consistent with a smooth, well-resolved solution in the chosen spectral basis.
%
\section{Training}\label{S4}
%
    In practice, the code runs transparently on either CPU or GPU; our benchmarks were obtained on NVIDIA P100 GPUs or a Ryzen 7 7840HS–class CPU (16 threads). On both hardware configurations, typical runs require $\mathcal{O}(1$–$3)$ minutes to reach a residual loss $\mathcal{L} \simeq 10^{-4}$, and about $\sim 20$ minutes to reach $\mathcal{L} \simeq 10^{-5}$ where it tends to stabilize\footnote{Because the governing equations enter directly in the loss, PINNs and \texttt{SpectralPINN} do not suffer from issues common in standard ML networks, such as overfitting, poor generalization, and reliance on large labeled datasets, as the governing physics directly constrains the solution space.}.
%
    \subsection{Optimizer}\label{S4.1}
%
    A key ingredient in our training procedure is an optimizer that can handle complex-valued parameters in a stable way. Many of the trainable quantities in our models (\textit{e.g.,} the spectral amplitudes and eigenvalues) are complex, and naively applying the standard real-valued \texttt{Adam} optimizer to their real and imaginary parts can lead to sub-optimal behavior. We therefore implement a custom first-order optimizer, \texttt{HybridAdamDecoupled} (\texttt{HAdamD}), which generalizes \texttt{Adam} to the complex domain while keeping a simple interface.
   
    For reference, in the purely real-valued case, with parameters $u_t \in \mathbb{R}^n$ and gradients $U_t = \nabla_u \mathcal{L}(u_t) \in \mathbb{R}^n$, \texttt{Adam} updates the biased first and second moments as
    \begin{equation}
        m_t = \beta_1 m_{t-1} + (1-\beta_1)\,U_t\ , \qquad v_t = \beta_2 v_{t-1} + (1-\beta_2)\,U_t^2\ ,
    \end{equation}
    and then applies the bias-corrected step,
    \begin{equation}
        \hat{m}_t = \frac{m_t}{1-\beta_1 ^t}\ , \qquad \hat{v}_t =\frac{v_t}{1-\beta_2 ^t}\ , \qquad u_{t+1}=u_t -\alpha \frac{\hat{m}_t}{\sqrt{\hat{v}_t+\epsilon}}\ ,
    \end{equation}
    with learning rate $\alpha$, decay coefficients $\beta _1$ and $\beta_2$, and numerical stabilizer $\epsilon$. 

    In our setting, a substantial fraction of the trainable weights are complex-valued. Applying real \texttt{Adam} directly would either require flattening complex variables into unstructured real-imaginary blocks, or using a single scalar second moment for $|u_t|$, which can couple the real and imaginary directions in an undesirable way. To avoid this, \texttt{HAdamD} treats complex parameters as 2D real vectors but preserves a genuinely complex representation for the first moment.

    Let $u_t \in \mathbb{C}^{n}$ be a vector of complex numbers with gradients $U_t \in \mathbb{C}^n$. We denote the real and imaginary parts as $\mathbb{R}$ and $\mathbb{I}$, respectively, such that $u_t$ and $U_t$ are written as
    \begin{equation}
        u_t = u_t ^{\mathbb{R}}+ i u_t ^{\mathbb{I}}\ , \qquad U_t = U_t ^{\mathbb{R}}+ i U_t ^{\mathbb{I}} \ .
    \end{equation}
    The optimizer maintains a complex first moment $m_t \in \mathbb{C}^n$, and two decoupled second moments: $v_t ^{\mathbb{R}}$, $v_t ^{\mathbb{I}}\in \mathbb{R}^2$, tracking the variance along the real and imaginary axes separately. The moment updates are
    \begin{equation}
        m_t = \beta_1 m_{t-1}+(1-\beta_1) U_t\ , \,\,\,\,\,\, v_t ^{\mathbb{R}} = \beta_2 v_{t-1} ^{\mathbb{R}}+(1-\beta_2)(U_t^{\mathbb{R}})^2\ ,\,\,\,\,\, v_t ^{\mathbb{I}} = \beta_2 v_{t-1} ^{\mathbb{I}}+(1-\beta_2)(U_t^{\mathbb{I}})^2\ .
    \end{equation}
    After standard \texttt{Adam} bias correction,
    \begin{equation}
        \hat{m}_t = \frac{m_t}{1-\beta_1 ^t}\ , \qquad \hat{v}_t ^{\mathbb{R}} =\frac{v_t^{\mathbb{R}}}{1-\beta_2 ^t}\ , \qquad \hat{v}_t ^{\mathbb{I}} =\frac{v_t^{\mathbb{I}}}{1-\beta_2 ^t}\ ,
    \end{equation}
    the updated is applied axis-wise, with separated RMS (root-mean-square of the gradients) normalization for the real and imaginary parts:
    \begin{equation}
        \Delta u_t ^{\mathbb{R}} = \frac{{\rm Re}(\hat{m}_t)}{\sqrt{\hat{v }_t ^{\mathbb{R} } }+\epsilon }\, \qquad \Delta u_t ^{\mathbb{I}} = \frac{{\rm Im}(\hat{m}_t)}{\sqrt{\hat{v}_t ^{\mathbb{I}}}+\epsilon}\ ,\qquad u_{t+1} = u_{t}-\alpha \big[\Delta u_t ^{\mathbb{R}}+i \Delta u_t ^{\mathbb{I}}\big]\ .
    \end{equation}
    Thus, the direction of descent is still governed by a single complex first moment $m_t$, but the adaptive scaling is decoupled per axis, providing distinctive learning rates along the real and imaginary directions. In practice, this improves stability for complex weights whose real and imaginary parts experience gradients of different magnitude or curvature.

    For purely real parameters, \texttt{HAdamD} reduces exactly to the standard \texttt{Adam}. In that case, we store only a real first moment and reuse a single second-moment buffer $v_t ^{\mathbb{R}}$, discarding the imaginary branch. As a result, a single optimized instance can seamlessly handle models with arbitrary mixtures of real and complex tensors, using standard \texttt{Adam} updates on real-valued layers and our decoupled complex variant where needed, without changes to the training loop.
%
    \subsection{Scheduler}\label{S4.2}
%
    The networks are trained with an alternating (``block-coordinate'') optimization scheme that separates the update of the field amplitudes from the update of the eigenpair ($\omega, \Lambda$). This proves much more stable than simultaneously updating all parameters with a single optimizer.

    We split the parameters into two groups: eigenpair block, $\Theta_{\omega\Lambda}\equiv \{\omega,\Lambda\}$; weights, $\Omega$, and bias, $b$ (PINN), or amplitudes, $\mathcal{A}$ (SpectralPINN), block, $\Theta_{\rm Amp} \equiv\{\mathcal{A},\, \Omega,\, b\}$. With each block assigned its own instance of the complex-aware \texttt{HAdamD} described above, one acting only on the eigenpair parameters and one acting only on the amplitude/weight parameters. This explicit splitting avoids pathological couplings between the stiff eigenvalue directions and the high-dimensional network's amplitudes/weights and biases, and empirically leads to much more stable and monotonic convergence.

    The training follows an alternating phase structure, where each outer training epoch consists of two successive inner phases:
    \begin{itemize}
        \item[1] \textbf{Amplitude relaxation} \textit{(Phase A)}. The eigenpair block, $\Theta_{\omega \Lambda}$, is frozen by requiring no gradient computation on $\omega$ and $\Lambda$, while gradients are enabled only for $\Theta_{\rm Amp}$. We then perform $K_U=10$ inner optimization steps of \texttt{HAdamD} on the function residual loss, $\mathcal{L}\big(\Theta_{\rm Amp}|\Theta_{\omega\Lambda}\big)$, using fresh computation graphs built from detached copies of the collocation mesh $(x,y)$. Intuitively, this phase ``relaxes'' the networks towards the current eigenpair, given fixed $\omega$ and $\Lambda$.
        \item[2] \textbf{Eigen-pair correction} \textit{(Phase B)}. Next, the roles are inverted: the amplitudes/weights and biases are frozen and only the $\omega$ and $\Lambda$ terms are allowed to update. A single optimizer step is taken on the same loss functional, now viewed as $\mathcal{L}(\Theta_{\omega \Lambda}\,|\,\Theta_{\mathrm{Amp}})$, which adjusts the eigen-frequency and separation constant to better satisfy the function's residual currently relaxed field. The resulting scheme is a simple block-coordinate descent: several inner steps on the amplitude followed by one outer step on the eigenpair, repeated until convergence.
    \end{itemize}
    This alternating pattern effectively decouples the ``shape'' update (field amplitudes) from the network update (eigenpair), and avoids the noisy competition that occurs when both are updated simultaneously with a single optimizer.

    In our case, both optimizers are equipped with independent plateau-based learning-rate schedulers. Concretely, we use a \texttt{ReduceLROnPlateau}-type scheduler that monitors the outer-epoch loss $\mathcal{L}_t$ and only decreases the learning rate when the optimization has stalled for $100$ epochs (\textit{aka} patience). After each outer-epoch $t$, the current loss is passed to the scheduler; if $\mathcal{L}_t$ fails to improve beyond a small relative tolerance over the defined patience window, the learning rate $\alpha_t$ is reduced by a constant factor $\eta=0.95$ as $\alpha_{t+1}^{(\cdot)}=\eta\, \alpha_t^{(\cdot)}$, where $^{(\cdot)}$ denotes either the amplitude block or the eigenpair block. In practice, this yields a piecewise constant ``cooling'' schedule: the learning rate stays high while the loss is still making progress, and is only decreased once the optimization enters a plateau. This behavior is particularly convenient for stiff eigenvalue problems, where early aggressive decay would prematurely freeze the frequency update, but persisting with a large step-size after convergence leads to noisy oscillations.

    We impose a lower bound on the learning rate, $\alpha_{min}$, so that repeated plateau events cannot drive the step-size to zero. Training is terminated once the amplitude learning rate reaches this floor (here we use $\alpha_{\mathrm{min}} = 10^{-14}$ for the amplitude block), which provides a simple and robust stopping criterion. Since we pre-train our networks with a previously converged solution, we initialize the learning rates as $\alpha^{\omega\Lambda} = 10^{-4}$ and $\alpha^{\mathrm{Amp}} = 10^{-3}$. This plateau-based schedule is slower than a simple exponential decay but allows us to reach significantly higher accuracy.
    \begin{table}[ht]
     \centering
     \caption{Percentual cumulative ($\mathbb{R}+\mathbb{I}$) relative error of the perturbation frequency, $\omega$, with respect to Leaver’s continued-fraction value for $(s,\ell,m)=(-2,2,0)$ and $a/M=0.90$ and hard normalization (see Sec.~\ref{S5}). We list the numerical schemes, the underlying equation (separated ODE vs full 2D PDE) and the incremental impact of the float type, optimizer, and scheduler. The Base configuration uses a standard \texttt{float64} type, the standard \texttt{Adam} optimizer, and a single \texttt{ReduceLROnPlateau} scheduler acting on all parameters.}
     \vspace{2mm}
     \label{T2}
        \begin{tabular}{cc|cccc}
            \hline
            Scheme & Eq. type & Base & +$\mathbb{C}$-dtype & +\texttt{HAdamD} & +Alt. Sched. \\
            \hline
            PINN & ODE & $0.138$ & $0.113$ & $0.069$ & $0.050$ \\
            SpectralPINN & ODE & $0.131$ & $0.126$ & $0.016$ & $0.002$ \\
            SpectralPINN & PDE & $0.142$ & $0.134$ & $0.058$ & $0.011$ \\
            \hline
        \end{tabular}
    \end{table}
    Table~\ref{T2} summarizes how each ingredient in our training stack affects the final QNM accuracy. Moving from the Base configuration to a complex dtype produces a modest improvement, while the complex-aware \texttt{HAdamD} optimizer yields a much larger reduction in the frequency error, especially for \texttt{SpectralPINN}. The alternating (block-coordinate) schedule provides the final order-of-magnitude gain for the spectral runs, bringing the ODE case down to the $\sim 10^{-5}$ level and the 2D PDE case to the $\sim 10^{-4}$ level in cumulative $(\mathbb{R}+\mathbb{I})$ relative error.
%
    \subsection{Normalization conditions}\label{S4.3}
%
    The Teukolsky QNM problem is a homogeneous linear eigenvalue problem. Once the eigenpair $(\omega,\Lambda)$ is fixed, the reduced radial and angular equations admit a scaling symmetry: if $(f,g)$ is a solution, then $(C f, C^{-1} g)$, where $C\in \mathbb{C}$ is a constant, yields the same physical perturbation and the same eigenvalues, because the Teukolsky operator is linear and only the product $f(r)\cdot g(\theta)$ enters the reconstructed field. The same reasoning applies for the 2D formulation: the joint field $p(x,y)$ satisfies a linear PDE and is therefore defined only up to an overall multiplicative constant, $p \rightarrow C\,p$. 
    
    This scaling freedom is physically irrelevant -- QNM amplitudes are fixed by the initial data rather than by the mode equation itself -- but it introduces a degeneracy in any optimization-based solver, since an entire one-parameter family of eigenfunctions corresponds to the same eigenvalue. For this reason one must supplement the Teukolsky equations with a normalization condition (NC) that fixes the overall scale of the mode. Following Leaver's original construction, we do so by prescribing the value of the reduced solutions at specific points in the compactified domain, which uniquely selects a representative within each equivalence class of rescaled eigenfunctions. Following Leaver, we choose the NC
    \begin{equation}
        f(1) = 1\ ,\quad  g(-1) = 1\ \qquad \vee \qquad \phi (1,-1) = 1\ .
    \end{equation}
    We have then two options to impose the NC. The first, and the easiest procedure is by considering a \textit{soft normalization}\footnote{In much of the PINN literature this is described as a soft/hard boundary condition. In our case the physical boundary conditions at the horizon and at infinity have already been imposed in a hard way through the analytic ans\"atze of Sec.~\ref{S2}; what we impose here is a point normalization of the eigenfunctions, so we refer to it simply as a normalization condition.}. A soft implementation consists in imposing the NC through the interactive solving process. In our case, the soft normalization is imposed through the loss by adding a term that penalizes the solution if it moves away from the Leaver normalization,
    \begin{equation}
        \mathcal{L} = \mathcal{L}_{\rm bulk}+\beta \mathcal{L}_{\rm NC}\ ,
    \end{equation}
    with $\mathcal{L}_{\rm NC}=|f(1)-1|+|g(-1)-1|$ or $\mathcal{L}_{\rm NC}=|\phi(1,-1)-1|$, depending on the function under study, and $\mathcal{L}_{\rm bulk}$ denoting the loss associated with the two ODEs or the single PDE residual. The dimensionless weight $\beta$ controls the relative importance of the normalization term in the total loss. We find that the choice of $\beta$ has a significant impact on the trained values of $(\omega,\, \Lambda)$: larger $\beta$ generally leads to better agreement with Leaver’s continued-fraction results. In what follows we consider two prescriptions: \textit{i)} a constant weight $\beta = N_x\cdot N_y$, which we refer to as the \emph{soft} normalization; and \textit{ii)} a dynamically increasing weight $\beta = \mathrm{epoch}$, which gradually tightens the normalization during training and will be referred to as the \emph{dynamical} normalization. 

    An alternative procedure to impose Leaver's normalization is through a \textit{hard enforcement} framework. Following the procedure presented in \cite{luna2023solving}, for the ODE case we choose,
    \begin{equation}\label{E4.3}
        f(x)=\big( e^{x-1}-1\big)\mathcal{F}(x)+1\ , \qquad \qquad g(y) = \big( e^{y+1}-1\big)\mathcal{G}(y)+1\ .
    \end{equation}
    Concerning the hard enforcement for the 2D PDE, one might naively attempt a direct analogue of \eqref{E4.3}, \emph{i.e.,} $p(x,y)=\big(e^{x-1}-1\big)\big(e^{y+1}-1\big)\,\mathcal{P}(x,y)+1$. However, this choice forces $p(1,y)=1$ and $p(x,-1)=1$ for all $x$ and $y$, effectively pinning both axes to the normalization value and significantly distorting the gradients near the boundaries. In practice, this over-constrains the Teukolsky PDE and shifts the trained $(\omega,\Lambda)$ away from the Leaver values and from the soft-normalized runs. A simpler hard normalization relies on a point normalization that keeps the same gradients as the original function,
    \begin{equation}\label{E4.4}
        P(x,y) = 1+\big[\mathcal{P}(x,y)-\mathcal{P}(1,-1)\big] \ .
    \end{equation}
    This leaves all spatial derivatives unchanged, $\partial_x \mathcal{P} = \partial_x p$ and $\partial_y \mathcal{P} = \partial_y p$, and therefore preserves the derivative structure of the 2D field while fixing its overall amplitude. Thereinafter, we refer to solutions obtained with \eqref{E4.3} and \eqref{E4.4} as hard. The point normalization at $(x,y)=(1,-1)$ in the 2D formulation is chosen precisely so that, when the solution is separable, the effective normalization of the radial and angular factors matches Leaver's convention $f(1)=g(-1)=1$. In practice, these hard normalizations are implemented through analytic masks multiplying the network outputs. For the separated ODEs we write the trainable functions as
    \begin{equation}
        f(x) = 1 + M_f(x)\,\mathcal{F}(x)\ , \qquad g(y) = 1 + M_g(y)\,\mathcal{G}(y)\ ,
    \end{equation}
    with mask functions $M_f(x)=e^{x-1}-1$ and $M_g(y)=e^{y+1}-1$ that vanish exactly at the normalization points, $M_f(1)=0$ and $M_g(-1)=0$. The networks therefore only learns the auxiliary functions $\mathcal{F}$ and $\mathcal{G}$, while the analytic masks guarantee $f(1)=g(-1)=1$ for any value of the network weights, \textit{i.e.} the Leaver normalization is enforced identically at every training step. Crucially, we also exploit the analytic form of the masks to compute the derivatives of $f$ and $g$ explicitly. Denoting derivatives with respect to the compactified coordinate by primes, we use the product rule
    \begin{equation}
        f'(x) = M_f'(x)\,\mathcal{F}(x) + M_f(x)\,\mathcal{F}'(x)\ , \qquad
        f''(x) = M_f''(x)\,\mathcal{F}(x) + 2 M_f'(x)\,\mathcal{F}'(x) + M_f(x)\,\mathcal{F}''(x)\ ,        
    \end{equation}
    and analogously for $g(y)$. The terms $\mathcal{F}'$, $\mathcal{F}''$, $\mathcal{G}'$ and $\mathcal{G}''$,  are obtained from the trained network, while $M_f$, $M_f'$, $M_f''$, $M_g$, $M_g'$, and $M_g''$ are known analytic functions evaluated once on the collocation nodes. As a result, all spatial derivatives entering the Teukolsky operator are given by closed-form expressions, and we avoid relying on automatic differentiation to construct second-order derivatives of the network output, which will be completely absent while using \texttt{SpectralPINN}. This improves both numerical accuracy and efficiency of the solver. On the training side, due to the hard enforcement scheme the resulting loss function lacks the normalization term, and can be simply casted as $\mathcal{L} = \mathcal{L}_{bulk}$ so that the loss is just a norm of the Teukolsky residual evaluated at the collocation points.
    \begin{figure}[h!]
	   \centering
	   \includegraphics[width=0.7\textwidth]{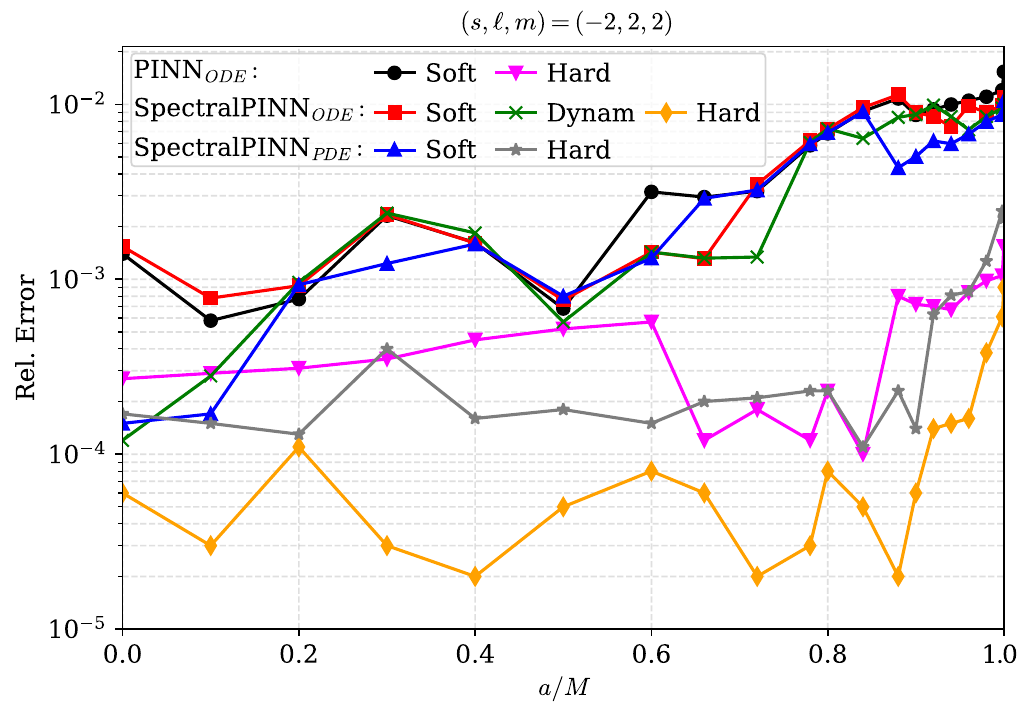}\\
	   \caption{Cumulative ($\mathbb{R}+\mathbb{I}$) relative error of the perturbation frequency $\omega$ with respect to Leaver’s continued-fraction values, as a function of $a/M$ with $N_x = N_y = 100$. Results are shown for separated-ODE PINNs with soft (black circles) and hard (magenta inverted triangles) normalization; separated-ODE SpectralPINN runs with soft (red squares), dynamical (green crosses) and hard (orange diamonds) boundary conditions for $N=L=25$; and for the full 2D PDE SpectralPINN with soft (blue triangles) and hard (grey stars) normalization for $N=L=25$.}
	   \label{F2}
    \end{figure}	

    Fig.~\ref{F2} shows the cumulative $(\mathbb{R}+\mathbb{I})$ relative error of the fundamental gravitational mode $(\ell,m)=(2,2)$ as a function of the spin $a/M$ for all solvers and normalization schemes. For the baseline PINN$_{\text{ODE}}$, soft normalization yields errors at the $10^{-3}$–$10^{-2}$ level, which grow towards high spins, while hard normalization improves the accuracy by a factor of a few but still saturates around $\sim 10^{-3}$. Switching to \texttt{SpectralPINN} systematically reduces the error: in the separated case, the hard normalization scheme achieves the best performance, with errors in the range $10^{-5}$–$10^{-4}$ at low/intermediate spins and remaining below $\sim 5\times 10^{-4}$ up to $a/M\simeq 0.95$, whereas the soft and dynamical NC variants stay around $10^{-3}$. Solving the full 2D PDE is naturally slightly less accurate, but the hard \texttt{SpectralPINN}$_{\text{PDE}}$ run still tracks Leaver’s frequencies within $(1$–$3)\times 10^{-4}$ across the spin range, and the soft variant remains at the $\sim 10^{-3}$ level. Overall, hard Leaver normalization combined with the spectral architecture yields the most accurate and robust QNM frequencies, while the loss of precision when moving from the separated ODE system to the full 2D PDE remains modest. We also observe an overall difficulty, and consequent lack of accuracy as $a/M \to 1.0$, which in the \texttt{SpectralPINN} case may be attenuated by using more basis terms. We would also like to point out that, due to our complex optimizer, while not explicitly shown here, the two components of the frequency ($\mathbb{R}$ and $\mathbb{I}$) equally contribute to the cumulative error.
    \begin{figure}[h!]
	   \centering
	   \includegraphics[scale=0.6]{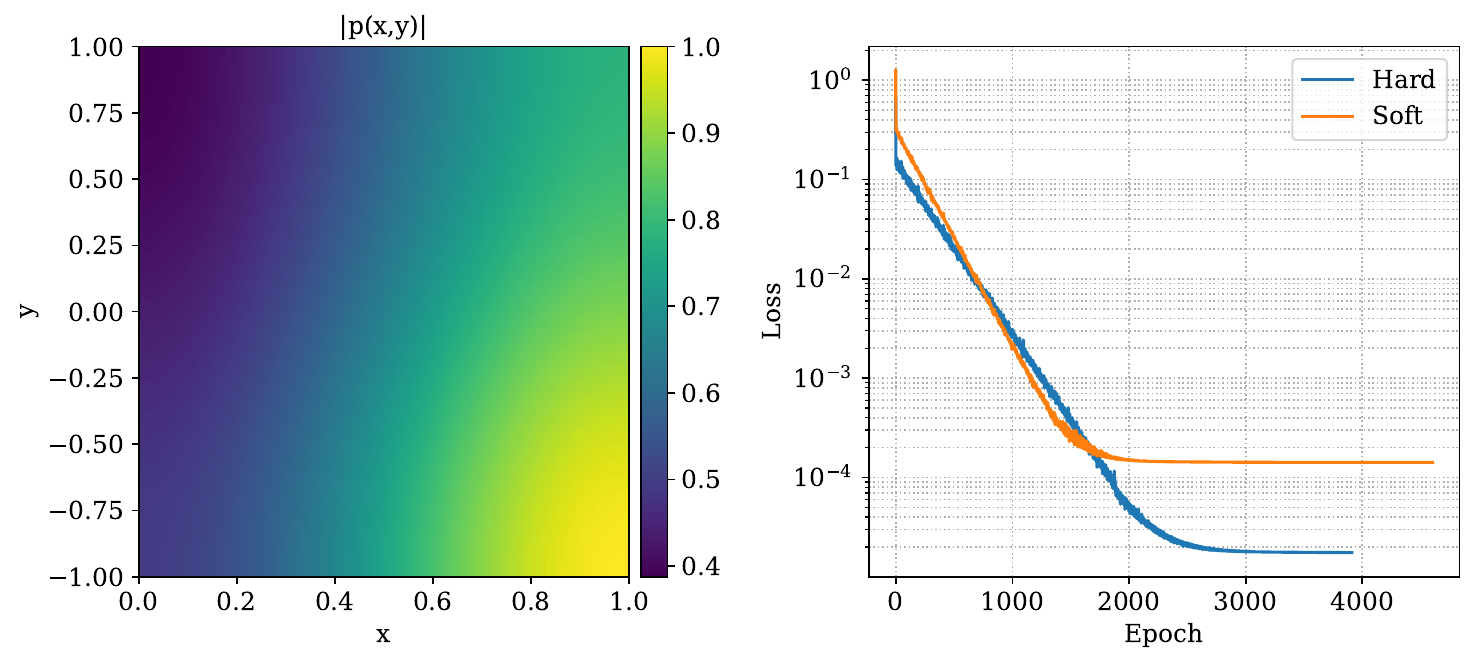}\\
	   \caption{Modulus of the 2D solution $|p(x,y)|$ (left) and training loss as a function of epoch for soft and hard NC (right).}
	   \label{F3}
    \end{figure}	

    Fig.~\ref{F3} illustrates the behavior of the full 2D solution for the fundamental gravitational mode. The left panel shows the modulus $|p(x,y)|$ over the compactified domain, for a representative spin value, displaying a smooth profile that remains regular at the boundaries and peaks near the outer edge of the radial coordinate ($x=1$ near horizon), as expected from the ringdown-dominated region outside the horizon. The right panel compares the training histories for the soft and hard NC schemes in the 2D \texttt{SpectralPINN}. Both runs exhibit an initial rapid decay of the loss over several orders of magnitude, but the soft-normalized model plateaus at a loss of $\mathcal{O}(10^{-4})$, whereas the hard-normalized run continues to decrease and ultimately settles at a loss of a few $\times 10^{-5}$. This confirms that enforcing Leaver’s normalization analytically in the 2D formulation not only lowers the final residual, but also yields a smoother and more robust optimization trajectory.
%
\section{Proof of concept}\label{S5}
%
    In this section, we illustrate how \texttt{SpectralPINN} can be used as a practical tool to probe non-Kerr deviations in the ringdown regime. A key motivation for developing the 2D \texttt{SpectralPINN} formulation is precisely its ability to treat perturbations of Kerr that spoil the exact separability of the Teukolsky equation. Many proposed deviations from Kerr -- including Kerr BHs with bosonic hair~\cite{HerdeiroRadu2015,HerdeiroRaduRunarsson2016,herdeiro2014kerr,delgado2021kerr} -- lead to effective wave operators with explicit $r\,$--$\,\theta$ couplings and are no longer reducible to a pair of decoupled radial and angular ODEs. In such cases, standard tools based on Leaver's continued fractions or purely separated spectral methods can only be applied at the cost of solving large coupled systems in a multipolar basis, whereas the 2D \texttt{SpectralPINN} can be used with minimal changes as long as the PDE remains linear and the boundary conditions retain the same asymptotic structure. 
    
    Here we focus on a controlled ``toy'' deformation of the Kerr Teukolsky operator that introduces a small, quadrupolar, non-separable perturbation while preserving the asymptotics. We first show that the 2D \texttt{SpectralPINN} can robustly track the resulting QNM branch as a function of the deformation parameter, and then map the induced shifts in the fundamental gravitational modes $(\ell,m)=(2,0)$ and $(2,2)$ into time-domain signals. Finally, by filtering these deformed ringdowns through the ET-D noise curve, we estimate the level at which the Einstein Telescope could detect or constrain such departures from the Kerr spectrum.
%
    \subsection{Quadrupolar non-separable deformation of Kerr}\label{S5.1}
%
    To mimic this situation in a controlled setting, we consider a simple quadrupolar deformation of the Kerr Teukolsky operator, which in the hard NC \texttt{SpectralPINN} case is equivalent to writing the loss function
    \begin{equation}\label{E5.1}
        \mathcal{L}_\varepsilon = \mathcal{L}_{\rm Teuk} + \varepsilon\,V(x,y)\,p(x,y)\ , \qquad \qquad V(x,y) = x^{3} P_2(y)\,,
    \end{equation}
    where $P_2(y) = \frac{1}{2}\big( 3y^2-1\big)$ is the Legendre polynomial associated with a mass quadrupole and $\varepsilon$ is a small, dimensionless deformation parameter. In the physical variables this corresponds to an effective quadrupolar correction $\propto P_2(\cos\theta)/r^3$, analogous to the leading multipolar deviation $\delta Q\,P_2(\cos\theta)/r^3$ that appears in quasi-Kerr or bumpy-BH parametrization. The term $V(x,y)p$ preserves linearity and the asymptotic structure of the problem: it vanishes as $r\to\infty$ ($x\to0$) and remains finite near the horizon. However, it explicitly couples the compactified radial and angular coordinates and therefore breaks exact separability.

\begin{figure}[h!]
	   \centering
	   \includegraphics[scale=0.6]{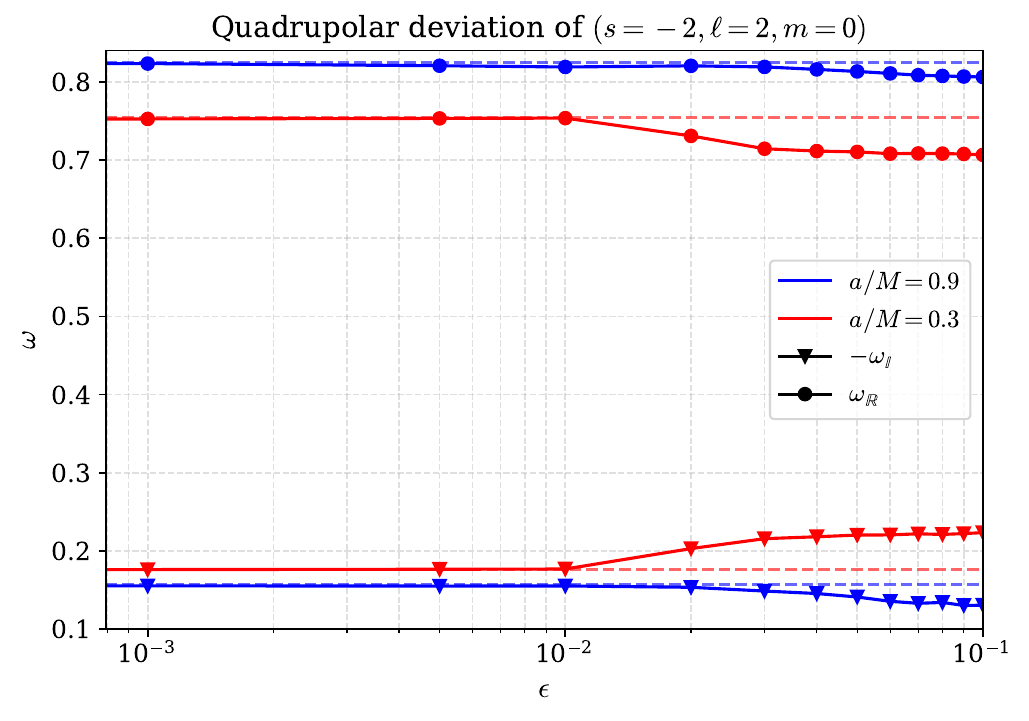}
	   \caption{Dependence of the fundamental $s=-2,\,\ell=2,\, m=0$ QNM frequency on the quadrupolar deformation parameter $\omega(\varepsilon)$ for $a/M=\{0.3,\, 0.9\}$. The dashed lines correspond to the theoretical expected value: $\omega = 0.7540-i\, 0.1767$ for $a/M=0.3$ (red) and $\omega = 0.8240-i\, 0.1570$ for $a/M = 0.9$ (blue).}
	   \label{F4}
    \end{figure}

    In the results below, Fig.~\ref{F4}, we take the converged Kerr solution at $\varepsilon=0$ for a representative configuration (\textit{e.g.,} the fundamental $s=-2,\,\ell=2,\,m=0$ mode at $a/M=0.9$) and use the 2D \texttt{SpectralPINN} to solve the deformed Teukolsky equation for a series of $\epsilon$ values. For each $\varepsilon$ we minimize $\mathcal{L}_\varepsilon$ until the residual stagnates $\sim 10^{-4}$, and we extract the corresponding eigen-frequency $\omega(\varepsilon)$ and 2D eigenfunction $p_\varepsilon(x,y)$. 

    Fig.~\ref{F4} shows how the real and imaginary parts of the fundamental $(\ell,m)=(2,0)$ QNM frequency respond to the quadrupolar deformation. For very small deformations, $\epsilon \lesssim 10^{-2}$, both $\omega_{\mathbb{R}}$ (circles) and $\omega_{\mathbb{I}}$ (triangles) lie essentially on top of their Kerr reference values (horizontal dashed lines), confirming that the non-separable perturbation leaves the spectrum almost unchanged in this regime. As $\epsilon$ grows towards $\sim 10^{-1}$ the mode drifts smoothly away from the Kerr point. Even for the largest deformations shown the branch remains continuous and well behaved, with no indication of mode crossings or instabilities, which supports the use of a perturbative treatment when mapping these frequency shifts into ringdown mismatches in the following subsection.
%
    \subsection{Comparison with Einstein Telescope templates}\label{S5.1}
%
    Let us now estimate how well the ET could constrain departures from the Kerr ringdown spectrum~\cite{Punturo2010ET,Hild:2010id} as a function of the deviation parameter $\epsilon$\footnote{This follows the standard parameterized ringdown tests of the no-hair theorem developed in \cite{Berti2006,Gossan:2011ha,Meidam:2014jpa}.}. Deviations from Kerr are encoded in the QNM frequency
    \begin{equation}
        \hat{\omega}(\epsilon;a/M) = \hat{\omega}_\mathbb{R}(\epsilon;a/M)- i\,\hat{\omega}_\mathbb{I}(\epsilon;a/M)\ ,
    \end{equation}
    expressed in units of $1/M$. For each spin $a/M$ our Teukolsky-based solver provides the map $\epsilon \mapsto \hat{\omega}(\epsilon;a/M)$. We then convert to detector-frame quantities by fixing a source-frame mass $M$ (we take $M=70\,M_\odot$) and we define $\omega^{\rm phys}(\epsilon) = \hat{\omega}(\epsilon)\,c^3/(GM)$, so that
    \begin{equation}
        f(\epsilon) = \frac{\omega^{\rm phys}_\mathbb{R}(\epsilon)}{2\pi}\ ,\qquad \qquad \tau(\epsilon) = \frac{1}{ \omega^{\rm phys}_\mathbb{I}(\epsilon)}\ ,
    \end{equation}
    where $f(\epsilon)$ and $2\pi\,\tau(\epsilon)$ represent the physical frequency in Hertz and the characteristic damping time. In this work we focus on the gravitational modes with $s=-2,\,\ell=2,\,m=0$ and $m=2$, but the analysis is identical for any chosen QNM.

    The time-domain ringdown signal is modeled as a single, linearly polarized, damped sinusoid,
    \begin{equation}
        h(t;\epsilon) = A\,e^{-t/\tau(\epsilon)} \cos\bigl[2\pi f(\epsilon)\,t + \phi_0\bigr]\ ,
    \end{equation}
    defined on a finite window $0 \leqslant t \leqslant T$, sampled at frequency $f_s$ (so $t_k=k/f_s$ and $h_k(\epsilon)\equiv h(t_k;\epsilon)$). The overall amplitude $A$ and phase $\phi_0$ are kept fixed; all $\epsilon$–dependence enters through $f(\epsilon)$ and $\tau(\epsilon)$. We denote the Kerr reference waveform by $h_0(t)\equiv h(t;\epsilon=0)$.

    To account for the detector response we adopt the ET-D design sensitivity and model ET as a single effective detector characterized by the official \emph{sum} amplitude spectral density $S_h^{\rm (sum)}(f)$~\cite{Punturo2010ET,Hild:2010id}. Squaring this yields the one-sided noise power spectral density $S_n(f) = [S_h^{\rm (sum)}(f)]^2$, which we use to define the standard noise-weighted inner product $(h_1|h_2)$ and optimal SNR $\rho^2(h)=(h|h)$, as in \cite{Lindblom:2008cm}. We restrict the integration to a suitable band $f_{\min} \le f \le f_{\max}$ around the mode (\textit{e.g.,}\ $5\,{\rm Hz} \le f \le 2048\,{\rm Hz}$).

    To quantify how distinguishable a deformed signal $h_\epsilon$ is from the Kerr reference $h_0$, we introduce the normalized match
    \begin{equation}
        \mathcal{M}(\epsilon; t_c) = \frac{(h_0|h_\epsilon^{(t_c)})}{\sqrt{(h_0|h_0)\,h_\epsilon^{(t_c)}|h_\epsilon^{(t_c)})}}\ ,
    \end{equation}
    where $h_\epsilon^{(t_c)}(t)=h_\epsilon(t-t_c)$ and $t_c$ is a coalescence-time shift. We maximize over a small grid of $t_c$ around zero and define
    \begin{equation}
        F(\epsilon) = \max_{t_c}\, \mathcal{M}(\epsilon; t_c)\ ,  \qquad \qquad \mathcal{M}_{\rm mis}(\epsilon) \equiv 1 - F(\epsilon)\ .
    \end{equation}
    with $\mathcal{M}_{\rm mis}$ quantifying the mismatch. For small deformations this is well described by
    \begin{equation}
        \mathcal{M}_{\rm mis}(\epsilon) \simeq \alpha\,\epsilon^2 \qquad (|\epsilon|\ll 1)\ ,
    \end{equation}
    with $\alpha$ obtained by fitting $\mathcal{M}_{\rm mis}(\epsilon)/\epsilon^2$ over a set of small, non-zero $\epsilon$ values.

\begin{figure}[h!]
	   \centering
	   \includegraphics[scale=0.52]{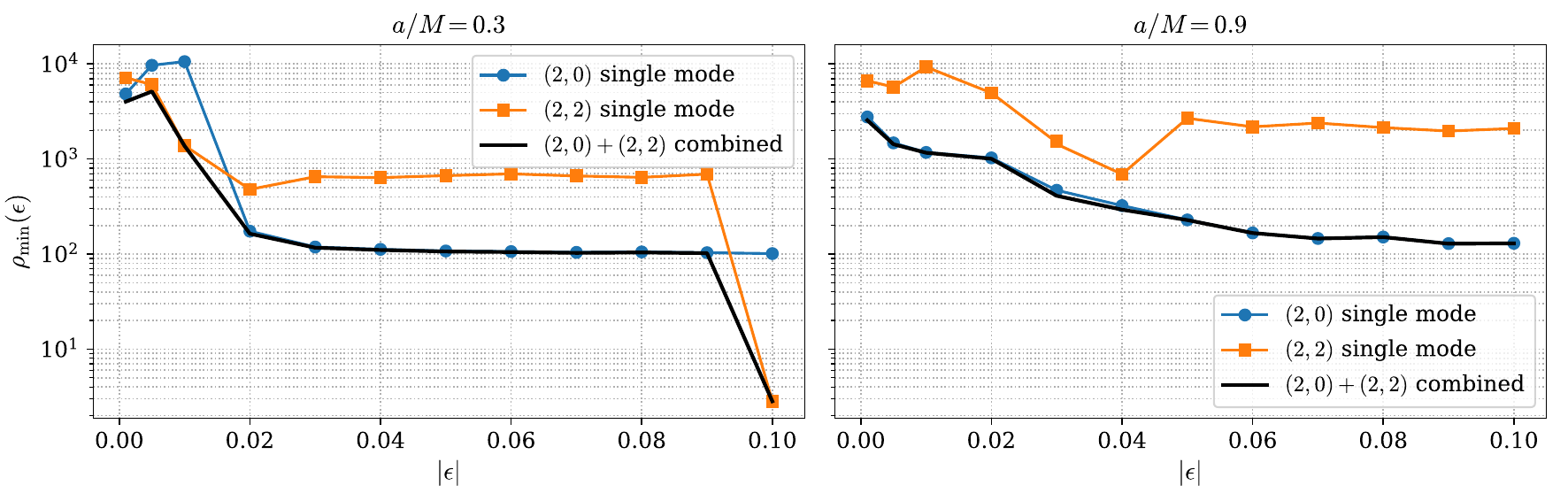}
	   \caption{Minimal per-mode ringdown SNR $\rho_{\min}(\epsilon)$ required to distinguish a deviation $\epsilon$ from Kerr at the $D=4$ level, using the ET-D noise curve. We show the single-mode thresholds for the $(2,0)$ and $(2,2)$ modes and the combined constraint when both modes are used jointly, for spins $a/M=0.3$ (left) and $a/M=0.9$ (right).}
	   \label{F5}
    \end{figure}

    Let $\rho^2=(h_0|h_0)$ be the SNR of the Kerr waveform and $\Delta h(\epsilon)\equiv h_\epsilon-h_0$ the waveform difference. In the small-$\epsilon$ regime one finds~\cite{Lindblom:2008cm}
    \begin{equation}
        (\Delta h|\Delta h)  \simeq 2\,\rho^2\,\mathcal{M}_{\rm mis}(\epsilon)\ .
    \end{equation}
    Imposing a distinguishability threshold $(\Delta h|\Delta h)\gtrsim D^2$ (with $D=4$) gives the minimum SNR required to resolve a deformation of size $\epsilon$,
    \begin{equation}
        \rho_{\min}(\epsilon) = \frac{D}{\sqrt{2\,\mathcal{M}_{\rm mis}(\epsilon)}}\ .
    \end{equation}
    Conversely, for an ET event with measured SNR $\rho_{\rm event}$ in the mode under consideration, the largest deformation compatible with a Kerr ringdown, $|\epsilon|_{\max}(\rho_{\rm event})$, is defined by $\rho_{\rm event}=\rho_{\min}(\epsilon)$. In the quadratic regime this reduces to
    \begin{equation}
        \rho_{\min}(\epsilon) \simeq \frac{D}{\sqrt{2\alpha}\,|\epsilon|}
  \quad\Rightarrow\quad |\epsilon|_{\max}(\rho_{\rm event}) \simeq \frac{D}{\sqrt{2\alpha}\,\rho_{\rm event}}\ ,
    \end{equation}
    making explicit that the constraint on $\epsilon$ improves inversely with the ringdown SNR. The same pipeline can be applied to any spin $a/M$ and any QNM family, and more realistic multi-mode or inspiral-merger-ringdown analyses can be built on top of this framework~\cite{Berti2006,Gossan:2011ha,Meidam:2014jpa}. Fig.~\ref{F5} recasts the ET-weighted mismatches into the minimal per-mode SNR $\rho_{\min}(\epsilon)$ required to distinguish a given deviation $\epsilon$ from the Kerr spectrum. For both spins, the curves diverge as $|\epsilon|\to 0$, reflecting the fact that an arbitrarily loud event would be needed to resolve an arbitrarily small deviation.
    As $|\epsilon|$ increases, $\rho_{\min}(\epsilon)$ decreases roughly as $1/|\epsilon|$, so that moderate deviations become accessible with realistic ringdown SNRs. At fixed $|\epsilon|$, the $(\ell,m)=(2,2)$ curves lie systematically below the $(2,0)$ ones, indicating that the $(2,2)$ mode is intrinsically more sensitive to the deformation in our toy model. The thick black line in each panel corresponds to the two-mode analysis, where the $(2,0)$ and $(2,2)$ modes are used jointly; this combined curve always lies below the single-mode thresholds, illustrating how multi-mode ringdown spectroscopy reduces the SNR requirement per mode. The corresponding constraints on the deviation parameter are summarized in Table~\ref{T3}. These numbers confirm the qualitative behavior seen in Fig.~\ref{F5}: the $(2,2)$ mode dominantly constrains the deformation, especially at low spin, while the inclusion of the subdominant $(2,0)$ mode provides a modest but consistent improvement in the high-spin case and leaves the low-spin constraint essentially unchanged.
    \begin{table}
        \centering
        \caption{Estimated upper bounds on the deviation parameter $|\epsilon|_{\max}$ from ET ringdown observations for two spins and different mode combinations. Single-mode bounds use only the indicated $(\ell,m)$ mode, while the combined rows assume a common deviation parameter for the $(2,0)$ and $(2,2)$ modes.}
        \label{T3}
        \vspace{2mm}
        \begin{tabular}{ccc}
            $a/M$ & Modes & $|\epsilon|_{\max}$ \\
            \hline
            0.9 & $(2,0)$             & 0.149 \\
            0.9 & $(2,2)$             & 0.074 \\
            0.3 & $(2,0)$             & 0.114 \\
            0.3 & $(2,2)$             & 0.020 \\
            \hline
            0.9 & $(2,0)+(2,2)$       & 0.066 \\
            0.3 & $(2,0)+(2,2)$       & 0.019 \\
        \end{tabular}
    \end{table}

%
\section{Conclusion}\label{S8}
%
    In this work, we introduced \texttt{SpectralPINN}, a hybrid PSM/PINN solver designed to compute Kerr QNMs by solving the Teukolsky equation in both its separated (radial/angular) form and its full 2D formulation. The method replaces standard neural activations with Chebyshev polynomials of the first kind, endowing the network with a spectral inductive bias while retaining the flexibility of PINN training. Leaver's normalization condition can be imposed either softly through loss penalties or exactly through analytic masks, allowing the solver to remain robust across a wide range of regimes.

    Benchmarking against Leaver’s continued–fraction approach~\cite{leaver1985analytic} demonstrates that the separated \texttt{SpectralPINN} with hard normalization reaches cumulative (real+imaginary) errors on the frequency at the $\sim 0.001\%$ level, while the 2D solver achieves $\sim 0.1\%$ accuracy across the whole range of spin parameter $a$. Crucially, these results were obtained without relying on exact separability of the Teukolsky operator or on analytic asymptotics beyond the horizon or at infinity. This establishes \texttt{SpectralPINN} as a solver that is simultaneously \emph{high-precision} and \emph{structurally flexible}: it can reach accuracies comparable to specialized spectral tools where separability holds, while continuing to function when separability is weakly or fully broken.

    To illustrate this capability, we applied the 2D SpectralPINN to a controlled quadrupolar deformation of the Teukolsky operator that couples the radial and angular coordinates and violates separability. The solver successfully recovered the deformed QNM spectrum and tracked continuous frequency shifts as a function of the perturbation amplitude. Mapping these shifts into time-domain ringdown waveforms and comparing them against the Einstein Telescope sensitivity curve~\cite{Punturo2010ET,Maggiore2020} showed that even small non-Kerr deviations could be constrained with realistic ringdown signal-to-noise ratios expected from next-generation detectors. This proof of concept highlights the potential of \texttt{SpectralPINN} as a practical tool for beyond-Kerr spectroscopy.

    Looking forward, the most compelling applications of the 2D solver lie in systems where separability is not guaranteed: Kerr BHs with bosonic hair (\textit{e.g.,}~\cite{HerdeiroRadu2015,HerdeiroRaduRunarsson2016,herdeiro2014kerr,delgado2021kerr}), horizonless COs~\cite{brito2016proca,schunck1996rotating,delgado2020rotating}, and modified gravity scenarios with additional degrees of freedom and different boundary conditions. These systems typically require large, coupled multipolar expansions when treated with continued fractions or traditional spectral decompositions, whereas the present approach can be extended to accommodate these modifications, provided the governing PDE remains linear and the asymptotics are under control. 

    Work is already underway to apply \texttt{SpectralPINN} to the computation of QNMs in hairy BHs, with the goal of providing mode catalogues and quantitative forecasts for spectroscopic tests of gravity in the era of Einstein Telescope, Cosmic Explorer, or LISA. The precision and adaptability of our method make it a natural building block for the next generation of theoretical predictions needed to maximize the physics return of forthcoming gravitational-wave observations.
%
\acknowledgments
    We would like to thank Ignacy Sawicki, Raimon Luna, José Blázquez-Salcedo and Guilherme Simplicio for their valuable discussions and comments. A. M. Pombo is supported by the Czech Grant Agency (GA\^CR) project PreCOG (Grant No. 24-10780S). We acknowledge the support of the European Consortium for Astroparticle Theory in the form of an Exchange Travel Grant. This study is supported by the Italian Ministry for Research and University (MUR) under Grant 'Progetto Dipartimenti di Eccellenza 2023-2027' (BiCoQ). 
%
%
\appendix

\bibliographystyle{JHEP}
\bibliography{biblio}

\appendix
%
\section{SpectralPINN weights}\label{A1}
%
\begin{table}[ht!]
  \centering
  \caption{First $6\times 6$ Chebyshev weights $\hat{a}_{ij}$ for $a/M=0.90$ and $(s,l,m)=(-2,2,0)$.}
  \label{tab:weights_a045_6x6}
  \scriptsize
  \setlength{\tabcolsep}{4pt}
  \begin{tabular}{c|cccccc}
    $i\backslash j$ & 0 & 1 & 2 & 3 & 4 & 5 \\
    \hline
    0 &
    8623.270+8666.060i &
    -1262.600-866.473i &
    75.272+33.049i &
    -3.674-0.792i &
    0.129-0.037i &
    0.013+0.020i \\
    1 &
    -13300.700-13382.300i &
    1946.690+1338.430i &
    -115.493-51.202i &
    5.282+1.250i &
    0.043+0.085i &
    -0.155-0.040i \\
    2 &
    3401.300+3460.230i &
    -498.629-346.695i &
    29.644+13.273i &
    -1.373-0.362i &
    -0.013+0.004i &
    0.056-0.000i \\
    3 &
    7681.340+7675.980i &
    -1122.860-766.306i &
    66.464+29.173i &
    -3.042-0.754i &
    0.013+0.019i &
    0.067-0.010i \\
    4 &
    -14821.500-14896.500i &
    2169.700+1489.230i &
    -129.166-56.803i &
    6.142+1.372i &
    -0.066+0.098i &
    -0.137-0.054i \\
    5 &
    15027.300+15178.400i &
    -2201.450-1519.140i &
    131.199+58.114i &
    -6.347-1.504i &
    0.161-0.021i &
    0.078+0.008i \\
  \end{tabular}
\end{table}
%

%
\section{Perturbation functions}\label{A2}
%
    Here we list the functions $P_{i,j}$ needed for the 2D PDE, \eqref{E2.16}.
    \begin{align}
     P _{0} &= 4\Bigg\{ \omega\Big[\mathrm{i}(3 + b) - 2 a m - a^{2} \omega + 2(1 + b) \omega\Big]+ x\Big[b - 2\mathrm{i}(1 + b) \omega + a^{2} \omega^{2} - 2 a m (\mathrm{i} + \omega)\Big] \nonumber\\
        &+ \frac{1}{2} x^{2}\Big[ -4 a^{3} m \omega- 2 a^{4} \omega^{2} + 2 a(1 + b)m(\mathrm{i} + 2\omega)  - (1 + b)(1 + 4\omega^{2})\nonumber \\
        &+ 2 a^{2}\big(2 + \omega\big(\mathrm{i}(1 + b) + 2(3 + b)\omega\big)\big)\Big]\Bigg\}(-1 + y^{2}) \nonumber \\
        &+  c\Bigg\{- 2(-1 + y^{2})|m+s|\big(1 + 2 a w(1 + y) + |m+s|\big) \nonumber \\
        &+4\Big[2 + m^{2} - 4 m y + 2 y^{2}+ a^{2} \omega^{2}(-1 + y^{2}) + 2 a \omega y(-1 + y^{2})\Big]  - (1 + y)^{2} |m+s|^{2} \nonumber\\
        &- 2\big(1 + 2 a \omega(-1 + y)\big)(-1 + y^{2})|m-s|- (-1 + y)^{2}|m-s|^{2} \Bigg\}\ ,\\
    P_{x} &= 4 c(-1 + y^{2})\Big[ x\big(-2 - (b + 2\mathrm{i} a m)u + 2 a^{2} x^{2}\big)+ 2\mathrm{i}\big(-1 + (1 - a^{2} + b)x^{2}\big)\omega\Big]\ , \\
    P_{xx} &= 4 x^{2} c^{2}(-1 + y^{2})\ ,\\
    P_{y} &= -4 c(-1 + y^{2})\Big( 2\big(y + a \omega(-1 + y^{2})\big)+ (1 + y)|m+s|+ (-1 + y)|m-s|\Big)\, \\
    P_{yy} &= -4 c(-1 + y^{2})^{2}.
\end{align}
with
    \begin{equation}
        b = \sqrt{1 - 4 a^{2}}\ , \qquad \qquad c = 1 + x(-1 + a^{2}x) \;=\; 1 - x + a^{2}x^{2}\ .
    \end{equation}
    In the same spirit as \cite{luna2023solving} we also list the functions $F_{i}$ and $G_j$ needed for the two ODEs \eqref{E2.12}.
    \begin{align}
        F &= \frac{1}{2}\bigg[2 \omega\Big(\mathrm{i}(3 + b) - 2 a m - a^{2} \omega + 2(1 + b) \omega\Big)+ 2\big(b - 2\mathrm{i}(1 + b) \omega + a^{2} \omega^{2} - 2 a m(\mathrm{i} + \omega)\big)x\nonumber\\
        & + \big( -4 a^{3} m \omega- 2 a^{4} \omega^{2}+ 2 a(1 + b)m(\mathrm{i} + 2\omega)\nonumber\\
        &- (1 + b)(1 + 4\omega^{2})+ 2 a^{2}\big(2 + \mathrm{i}(1 + b) \omega + 2(3 + b) \omega^{2}\big)\big)x^{2}\Big]\ ,\\
        F_\Lambda &=-\Lambda \,\big(1 + x(-1 + a^{2}x)\big)\ ,\\
        F_x &= \big(1 + x(-1 + a^{2}x)\big)\Big[x\big(-2 - (b + 2\mathrm{i} a m)x + 2 a^{2} x^{2}\big)+ 2\mathrm{i} \omega\big(-1 + (1 - a^{2} + b)x^{2}\big)\Big]\ ,\\
        F_{xx} &= x^{2}\big(1 + x(-1 + a^{2}x)\big)^{2}\ .
    \end{align}
    \begin{align}
        G &= 4\Big( 2 + m^{2} - 4 m y + 2 y^{2}  - 2 a y \omega + 2 a y^{3} \omega -a^{2} \omega^{2} + a^{2} y^{2} \omega^{2}\Big)- (1 + y)^{2}|m-s|^{2} \nonumber\\
        &- 2(-1 + y^{2})\big(1 + 2 a(-1 + y) \omega\big)|m+s|- (-1 + y)^{2}|m+s|^{2}\nonumber\\
        &- 2(-1 + y^{2})|m-s|\big(1 + 2 a(1 + y) |m+s|\big)\ ,\\
        G_\Lambda &= \Lambda(-1 + y^{2}) \ ,\\
        G_y &= -4(-1 + y^{2})\Big[ 2\big(y - a \omega + a y^{2} \omega\big) + (1 + y)|m-s|+ (-1 + y)|m+s|\Big]\ ,\\
        G_{yy} &= -4(-1 + y^{2})^{2}\ .
    \end{align}

\end{document}